\documentstyle[preprint,aps]{revtex}
\begin{document}

\draft

\title{Capacitance spectroscopy in quantum dots: Addition spectra
and decrease of tunneling rates}

\author{J. J. Palacios$^1$, L. Mart\'{\i}n-Moreno$^2$, G. Chiappe$^3$,
E. Louis$^3$, and C.Tejedor$^1$ }

\address{$^1$Departamento de F\'{\i}sica de la Materia Condensada.
Universidad Aut\'onoma de Madrid,
Cantoblanco, 28049, Madrid, Spain.  }

\address{$^2$Instituto de Ciencia de Materiales (CSIC).
Universidad Aut\'onoma de Madrid,
Cantoblanco, 28049, Madrid, Spain.  }

\address{$^3$Departamento de F\'{\i}sica Aplicada.
Universidad de Alicante,
Apartado 99, 03080 Alicante, Spain.  }

\date{\today}

\maketitle

\begin{abstract}
A theoretical study of single
electron capacitance spectroscopy in quantum dots is presented.
Exact diagonalizations and the unrestricted Hartree-Fock approximation
have been used to shed light over some of the unresolved aspects.
The addition spectra of up to 15 electrons is obtained and compared
with the experiment. We show evidence for understanding the decrease of
the single electron tunneling rates in terms of the behavior of
the $\omega \rightarrow 0$ spectral weight function.
\end{abstract}

Single electron capacitance spectroscopy (SECS) \cite{1,2}
has been a breakthrough in the experimental knowledge of the
electronic structure of a quantum dot (QD). Ashoori and
co-workers \cite{1,2} have been able to determine the energies required to
introduce electrons one by one, from 0 to 50, into a QD. The
electrons tunnel into the QD by means of a vertical gate bias, and
change the capacitance of the device. The measurement of that
capacitance as a function of the Fermi energy $E_F$ in one electrode shows a
discrete set of almost equally spaced peaks of different intensities.
A peak appears whenever $E_F=\mu(N)=E_0(N)-E_0(N-1)$, with $E_0(N)$ being the
ground state (GS) energy of $N$ electrons in the QD.
In this way, they were able to measure not only the
addition spectra as a function of $N$, but also
the tunneling rates of single electrons entering the QD when the
energy balance is fulfilled. The main output of the experiment
is to reveal a very rich dependence of $\mu (N)$
on an applied magnetic field $B$
as well as to show strong suppressions of the tunneling rates in
well-defined regions of the $B-N$ space.
The constant interaction (CI) model \cite{2} for the electron-electron
interaction cannot explain many of the features observed
in $\mu (N)$ vs. $B$,
the more serious drawback of the CI model being that a
single-particle picture cannot account for the observed suppression of the
tunneling rates in this experimental set up.

Theoretical attempts to go beyond the CI model and to
explain the properties of electrons in QD's run from the simplest
Thomas-Fermi \cite{maceuen}, Hartree \cite{jacobo} and Hartree-Fock \cite{wen}
descriptions of the electron-electron interaction to the more sophisticated
exact diagonalizations of the complete Hamiltonian \cite{4,5,7,8,9,10}.
So far, these "exact" calculations have been performed in two
regimes: either for small number of electrons (2 or 3), or restricted
to the Hilbert subspace of the two first Landau levels ,i.e., first Landau
index, spins up and down.
The aim of this work is the understanding of the SECS experimental results
which requires two ingredients lacking in previous theoretical attemps: i) the
analysis of $\mu $  and wavefunctions in a broad range of $B$ and $N$ and
ii) the study of peak intensities controlled by tunneling rates.

The Hamiltonian for interacting electrons is

\begin{equation}
H=\sum_{i,\sigma} \epsilon_{i,\sigma} c^{\dagger}_{i,\sigma}
c_{i,\sigma} + \frac{1}{2}
\sum_{i,j,k,l,\sigma,\sigma'}V_{ijkl}c^{\dagger}_{i,\sigma}
c^{\dagger}_{j,\sigma'}c_{k,\sigma'}c_{l,\sigma}
\label{mbham}
\end{equation}

where the sum runs over the single-particle states of a two-dimensional
parabolic potential defined in the $xy$ plane,
characterized by a frequency $\omega_0$, and embedded
in a magnetic field, perpendicular to the $xy$ plane.
$\epsilon_{i,\sigma}$ is the single-particle energy
of a state $i$ with spin $\sigma$
(an infinitesimal Zeeman term is included to remove degeneracies in
spin multiplets). These states are characterized by the Landau
index $n$ ($0 \rightarrow \infty$),
and by the third component of the angular momentum, $m$
($-n \rightarrow \infty$) \cite{10}.
$V_{ijkl}$ are the Coulomb interaction terms
given by four-dimensional integrals which include a cut-off ($\lambda_c$)
to account for the finite extension of the wave functions in the $z$ direction,
and $c^{\dagger}$ and $c$ are the creation and annihilation operators.
Throughout this work, the standard values of the effective mass and dielectric
constant in GaAs have been taken. $\omega_0$ is given by $\hbar\omega_0=5.4$
meV and $\lambda_c=10\AA$ (this choice is discussed below).

An exact diagonalization of the Hamiltonian up to 5 electrons is undertaken.
Due to the circular symmetry, the diagonalization can be
done in separate subspaces of configurations with the same
$z$-component of the total angular momentum, $M$, and
$z$-component of the total spin, $S_z$. The
lowest-lying eigenvalues and eigenfunctions in all the different
subspaces were obtained using standard diagonalization routines and
Lanczos-like procedures, and the GS is picked among all of them.
A relative error less than 0.1 \% for the energy in all the cases
was routinely achieved by including enough
$n$ and $m$ indices of the single-particle basis set:
Up to $n=1$ for high fields and $n=2$ below 1 Tesla.
The evolution of $\mu(N)$ with $B$ for up to $N=5$ is shown
in Fig. \ref{fig1} (solid lines).
For comparison, we show the results obtained with a basis set restricted to
the $n=0$ Landau index (dashed line) \cite{10}.
This case is much less time consuming than
the previous one, but significant deviations in the $\mu(N)$ plot
appear even at large values of $B$, as $N$ increases.
{}From now on, the discussion is based on the more-than-1-Landau-index
results.

Solid dots in Fig. \ref{fig1} mark the points at which
the complete polarization of the electrons in the
QD occurs. A small bump in $\mu(N)$ vs. $B$
is observed at those points, but at values of $B$ higher than those found
in the experiment of Ref. \cite{2}. There is an overall
disagreement between the scales
of Fig. \ref{fig1} and those of the data obtained in Ref. \cite{2}.
The confinement due
to the external potential and characterized by $\omega_0$
sets the curve $\mu(1)$ vs. $B$. Once this is fixed, the electron-electron
interaction determines the separation $\mu(2) - \mu(1)$ at $B=0$. Our
calculated
value is a factor of almost 2 too large compared with the experimental one.
This might indicate that the value of $\lambda_c$ chosen in the calculation of
$V_{ijkl}$ is too small, giving a too strong interaction.
On the other hand, the bump in $\mu(2)$ indicating the singlet-triplet
transition lies at a higher magnetic field than in the experimental data. One
should increase the interaction, i.e. shorten the cut-off, in order for this
transition to take place at lower fields (a realistic
Zeeman term does not affect
this transition noticeably). The only way of improving
both scales at the same time is to weaken
the strength of the interaction and to lower
the value of $\omega_0$: It is the dependence of $\mu(1)$ on $B$
that will be now in disagreement with the experiment.
All this seems to indicate that other
effects like non-parabolicity or disorder may be important \cite{matias}. For
simplicity's sake, we have chosen to work with the parabolic confinement that
gives us $\mu(1)$ in agreement with the experiment.
Bearing in mind the scale disagreements between theory and experiment,
this choice does not affect the main conclusions of our work.

For what follows, it is convenient to define a "bulk filling factor" $\nu$
in a QD as $\nu= \sum_{n,\sigma} \nu_{n,-n,\sigma}$, with
$\nu_{n,m,\sigma}$ being the single-particle state occupations.
This "bulk filling factor" reflects the Landau
levels occupation close to the center of the QD as $B \rightarrow \infty$,
and it is consistent with the usual definition of $\nu$ for
a two-dimensional system. With this convention for $\nu$, it is possible
to identify a phase diagram on Fig. \ref{fig1}.
The region to the left of the open dots corresponds to $\nu=2$, i.e.,
the electrons form a paramagnetic
compact droplet in the center of the QD, having $|S_z|$ minimum.
Defining $\nu_m=\sum_{n,\sigma} \nu_{n,m,\sigma}$, this region corresponds,
e.g., for $N=4$, to $\nu_0 \approx 2$, $\nu_1 \approx 2$,
$\nu_2 \approx 0$, etc.
A ferromagnetic state of $\nu=1$ occurs in the region to the right of the
solid dots, with the electrons forming a
compact droplet of $|S_z|$ maximum (for $N=4$: $\nu_0\approx 1$,
$\nu_1\approx 1$, $\nu_2 \approx 1$, $\nu_3 \approx 1$,
$\nu_4 \approx 0$ ...). At even higher magnetic fields, a $\nu < 1$
region appears \cite{10},
but it is not the aim of this paper to cover such a regime.

In the region delimited by the solid and open dots,
the transition from $\nu=2$ to $\nu=1$ or, in other words,
the paramagnetic-ferromagnetic transition (a process already considered
in references \cite{maceuen} and \cite{10}), takes place.
First of all, the GS "melts", i.e., the charge spreads
over a large number of single-particle states until a new compact state of
total spin $|S_z|+1$ is formed (one can say that a spin-flip process for an
electron has taken place). On increasing further the value of $B$,
the same process occurs for the remaining
spin-down electrons until complete polarization is achieved.
The last spin-flip process for $N=5$ is depicted in Figure \ref{fig2}.
It is in this region that the lowest-energy many-body
states of different subspaces
are very close in energy and cross each other continuously as $B$ changes.
The GS values of $M$ and $S_z$ change very
quickly with $B$ in the following way: $M$ always increases with $B$,
but $|S_z|$ has usually its
minimum possible value when the GS becomes one of the "melted" ones.
On the other hand, the GS's in regions with $\nu=2$ and $\nu=1$
are very stable in $B$ due to the large
separation in energy between the GS and the rest of lowest-lying
states of different subspaces. One can say that a "compact" QD develops in
these regions. The "melting" process here discussed is a typical
correlation effect that cannot be described by simple approaches as CI,
Hartree or Hartree-Fock in which the parameters $\nu _m$ can only assume
integer
values. Although a Thomas-Fermi approach does not have, in principle, such a
restriction, it has been used\cite{maceuen} assuming that the charge in each
Landau
level is quantized, something far from true in the "melted" regime.

As mentioned in the introduction, not only the fulfilment of the energy
balance determines the tunneling process. For temperatures and biases smaller
than the excitation energy (typically higher than 0.1 meV in our case)
in the QD, only GS's are essentially involved in the process, and
tunneling rates for single electrons are proportional to the spectral weight
at $\omega \rightarrow 0$, i.e., $E_F \rightarrow \mu^+(N)$ \cite{yo}, given by

\begin{equation}
\Delta(N)= \sum_{i,\sigma }
\langle \Phi_0^{(N)} | c_{i,\sigma }^\dagger | \Phi_0^{(N-1)} \rangle
\langle \Phi_0^{(N-1)} | c_{i,\sigma } | \Phi_0^{(N)} \rangle
\end{equation}

with $\Phi_0^{(N)}$ representing the $N$-electron many-body GS wave function.
Figure \ref{fig3} shows $\Delta(N)$  for $N=2,3,4$ and 5 as a function
of $B$. When the electrons form a compact system of $\nu=2$,
$\Delta(N)$ reaches a value close to 1. While the polarization
process is taking place in the QD either for $N$ or $N-1$,
$\Delta(N)$ is significantly reduced
as a consequence of the strong correlation in any of the $N$ or $N-1$ GS's.
Sometimes $\Delta(N)$ can even be zero due to the spin selection rule
$|S_z(N) - S_z(N-1)| > 1/2$ (in our calculation for
$B \simeq 4.8T$ for $N=4$ and $B \simeq 4.2$ and $5.4T$ for $N=5$),
It must be pointed out that, as $N$ increases, the
interval of decreased $\Delta(N)$ widens [see Fig. \ref{fig3}, for $N=3,4,5$].
This is due to the fact that the polarization requires a longer variation of
$B$
and, therefore, the presence of "melted" GS's becomes more probable for high
$N$.
As $B$ increases further and the ferromagnetic state of $\nu=1$ is reached
for both $N$ and $N-1$, the value of the
spectral weight again rises up to 1.  This phenomenon
has also been observed in the SECS experiment \cite{2}, although
for a large number of electrons in the QD: As $B$ increases
and the QD GS turns from the "compact" state of $\nu=2$ into that of
$\nu=1$, there is a region in which the capacitance signal reduces due to
the decrease of the tunneling rate.

A similar bump in $\mu(N)$ vs. $B$ and reduction of signal have been
experimentally observed
at lower values of $B$ and for a large $N$ \cite{2}.
The transition from $\nu=2$ to $\nu=1$ could be studied by exact
diagonalization, but this is not the case any more for a large number
of particles. In order to go beyond the small-$N$ regime,  we use
an unrestricted Hartree-Fock (UHF) approximation. The UHF approach
is undertaken by
substitution of pairs of electron operators by their mean-field
magnitudes $\langle c^{\dagger}_{i,\sigma'}c_{j,\sigma}\rangle$, which must
be calculated self-consistently\cite{luis}.

We have performed calculation similar to the one presented in Fig.
\ref{fig1}, with the same basis set used there, in the UHF approach.
The UHF GS's present a similar evolution with $B$ than those obtained
in the exact
calculation. At low field, a $\nu=2$ paramagnetic UHF GS appears. In this
case, the relative error of the UHF GS energy to the exact GS energy
is 6 \%, 4 \%, 3.2 \%, and 1.8 \% for $N=2,3,4$, and 5, respectively. At
high field, a $\nu=1$  ferromagnetic UHF GS develops. The relative error is
now 0.5 \%, 0.4 \%, 0.3 \%, and 0.2 \% for $N=2,3,4$, and 5, respectively. In
this approach, the transition from $\nu=2$ to $\nu=1$ takes place
by sudden spin flips of the spin-down electrons from the center to
the edge of the QD. The UHF approach
does always give integer values for $\nu_m$ and, therefore,
it cannot give the "melted" states which appear in the exact calculations.
Although the UHF approximation describes poorly the many-body
GS wave functions in the $\nu=2 \rightarrow \nu=1$ transition, the relative
error in the GS energy
remains below 6 \% even in the worst cases, and tends to decrease with
increasing $N$.  Figure \ref{fig4} shows the results of the UHF $\mu(N)$ for
up to 15 electrons with a basis set conveniently increased up to 64
single-particle states.
An approximate phase diagram can be traced on this plot:
(a) Solids dots follow the ferromagnetic transition,
i.e., the coming up of the $\nu=1$ QD (as can be seen in Fig. \ref{fig4},
it comes closer to that of the exact results as $N$ increases),
(b) open dots, the beginning of the polarization, i.e. the beginning of the
transition from $\nu=2$ to $\nu=1$, and (c) solid triangles,
the appearance of the $\nu=2$ QD
(this last transition appears also for $\mu(5)$ in the exact calculation).
As the number of electrons increases, a bump develops more
clearly (Fig. \ref{fig4}) and can be associated with the bump observed in
the SECS experiment \cite{2} at low fields and high number of particles.
In the transition from $\nu > 2$ to $\nu=2$
there is a new reduction of the capacitance signal in the experiment. This
fact can be understood if a decrease of the $\omega \rightarrow 0$
spectral weight takes place (as for the
$\nu=2 \rightarrow \nu=1$ transition), but, unfortunately,
the spectral weight is not well described within the UHF approach if
"melted" states are due to appear.
However, UHF describes properly the following experimental features:
(a) The monotonic behavior of $\mu(N)$ after the bump (due to the
presence of the stable "compact" $\nu=2$ state),
(b) the oscillating behavior in the $\nu=4$ to $\nu=2$ transition
(due to the progressive depopulation of higher Landau levels),
and (c) the anticrossings between successive traces
of $\mu(N)$ in this region. Some of these points are given by simple
models\cite{maceuen} while others are not, for instance point (c) cannot
be explained by a CI model.

A last comment remains to be done. If a finite Zeeman term is included,
the correlated GS's that appear in the polarization process of the QD and
carrying a small value of $|S_z|$, tend to be replaced by "compact" states
with higher values of $|S_z|$ (as in the UHF approach) \cite{10}.
However, with
a realistic $g$-Land\'e factor, and for a perpendicular magnetic field,
the results presented in this work do not change noticeably.

In this letter we have done a detailed study of the recent SECS experimental
data in QD's \cite{2}. The different regions appearing in the addition spectra
as the magnetic field changes have been obtained and analyzed by means of
exact diagonalizations and an UHF approach. The decrease of
the tunneling rates observed in the experiment is discussed in terms
of the reduction that the exact calculations show for the spectral weight
for $\omega \rightarrow 0$ when the QD becomes "melted".

We thank L. Brey and J. H. Oaknin for many useful discussions.
This work has been supported in part by the Comisi\'on
Interministerial
de Ciencia y Tecnolog\'{\i}a of Spain under contracts No. MAT 91-0905 and
MAT 91-0201 and by the Commission of the European Communities under
contract No. SSC-CT-90-0020.

\begin{figure}
\caption {$\mu(N)$ vs. magnetic field up to 5 electrons.
The dashed lines correspond
to the $n=0$ approach and the solid lines to the exact calculations.
The phase diagram delimited by the dots is discussed in the text.}
\label{fig1}
\end{figure}

\begin{figure}
\caption {Single-particle occupations of the GS for $N=5$ and
for three different values
of $B$: (a) Before the last spin-down electron leaves the center
of the dot ($B=5$ T), (b) when
the transit is taking place through a "melted" state ($B=5.5$ T), and (c)
when the final spin-polarized GS is reached ($B=6$ T).
As can be seen from the single-particle occupations in (b),
the intermediate "melted" state needs a high participation of
different configurations to be described.}
\label{fig2}
\end{figure}

\begin{figure}
\caption {$\omega \rightarrow 0$ spectral weight for $N=2,3,4$ and 5 as
a function of the magnetic field.}
\label{fig3}
\end{figure}

\begin{figure}
\caption {$\mu(N)$ vs. magnetic field up to 15 electrons in the UHF approach.
The phase diagram delimited by the symbols is discussed in the text.
The exact results for $\mu(N)$ are plotted in dotted lines as well as
the paramagnetic-ferromagnetic transition for this case (upward-pointing
arrows).}
\label{fig4}
\end{figure}

\end{document}